\documentclass[twocolumn,floatfix,showpacs,tightenlines,superscriptaddress]{revtex4-2}
\usepackage{mathtools}
\usepackage{bm}
\usepackage{dsfont,amsthm,amsbsy}
\usepackage{verbatim}
\usepackage{amssymb}
\usepackage{amsmath}
\usepackage{bbm}
\usepackage{graphicx}
\usepackage{epstopdf}
\usepackage{subfigure}
\usepackage{natbib}
\usepackage{epsfig}
\usepackage{amsfonts}
\usepackage{mathrsfs}
\usepackage{sidecap}
\usepackage{lipsum}
\usepackage[toc,page,title,titletoc,header]{appendix}
\usepackage[colorlinks,linkcolor=blue,citecolor=blue,anchorcolor=blue,urlcolor=blue]{hyperref}
\usepackage{hyperref}
\usepackage{resizegather}
\usepackage{tikz}
\usepackage{float}
\usepackage{mathbbol}
\usepackage[normalem]{ulem}
\usepackage{cancel}
\usepackage{upgreek}

\newcommand{\bl}{\begin{aligned}}
\newcommand{\el}{\end{aligned}}
\def\be{\begin{equation}}
\def\ee{\end{equation}}

\def\bi{\begin{itemize}}
\def\ei{\end{itemize}}
\def\bn{\begin{enumerate}}
\def\en{\end{enumerate}}
\def\bea{\begin{eqnarray}}
\def\eea{\end{eqnarray}}
\def\no{\nonumber}
\def\ba{\begin{array}}
\def\ea{\end{array}}
\def\bd{\begin{displaymath}}
\def\ed{\end{displaymath}}

\begin{document}

\title{Competition of long-range interactions and noise at ramped quench dynamical quantum phase transition:
The case of the long-range pairing Kitaev chain}

\author{R. Baghran}
\affiliation{Department of Physics, Institute for Advanced Studies in Basic Sciences (IASBS), Zanjan 45137-66731, Iran}

\author{R. Jafari}
\email[]{jafari@iasbs.ac.ir, raadmehr.jafari@gmail.com}
\affiliation{Department of Physics, Institute for Advanced Studies in Basic Sciences (IASBS), Zanjan 45137-66731, Iran}
\affiliation{School of Nano Science, Institute for Research in Fundamental Sciences (IPM), 19395-5531, Tehran, Iran}
\affiliation{Department of Physics, University of Gothenburg, SE 412 96 Gothenburg, Sweden}

\author{A. Langari}
\email[]{langari@sharif.edu}
\affiliation{Department of Physics, Sharif University of Technology, P.O.Box 11155-9161, Tehran, Iran}

\begin{abstract}
The nonequilibrium dynamics of long-range pairing Kitaev model with noiseless/noisy linear time dependent chemical potential, is
investigated in the frame work of dynamical quantum phase transitions (DQPTs). 
We have shown for the ramp crosses a single quantum critical point, while the short-range pairing Kitaev model displays a single
critical time scale, the long-range pairing induces a region with three DQPTs time scales. We have found that the region with three 
DQPTs time scales shrinks in the presence of the noise. In addition, we have uncovered for a quench crossess two 
critical points, the critical sweep velocity above which the DQPTs disappear, enhances by the long-range pairing exponent while decreases 
in the presence of the noise. On the basis of numerical simulations, we have shown that noise diminishes the long-range pairing inductions.
\end{abstract}

\maketitle

\section{Introduction}

Long-range interactions has been attracting great interest due to revealing surprising features 
\cite{Thouless1969,Fisher1972,Dutta2001,Luijten2001,Regemortel2016,Giuliano2018,Ares2018,Benito2014,Francica2022,Saikat2022,Dias2022,Simon2020,Vodola2014,Utso2018}.
Long-range systems are often fascinating approach to analyze the validity of the hypotheses 
that are otherwise clearly perceived for prototypical short-range systems. The remarkable experimental 
advancement in ultracold atomic platforms has triggered a plethora of theoretical studies and opened up 
the possibility of engineering and fine-tuning long-range systems with great accuracy 
\cite{schauss2012,britton2012,Islam2013,richerme2014n,Monroe2021,bloch2012}.

Moreover, the unprecedented advancement in recent years is impeccable enough to study the none-equilibrium dynamics of 
long-range systems in a controled manner \cite{Dupont2022,Langen2015,king2022,keesling2019}. Consequently, 
over the latest decades, theoretical and experimental research has raised a great deal of interest in 
non-equilibrium quantum phenomena \cite{Polkovnikov2011}, which has led to the discovery of some intriguing 
physics, including the observation of Kibble-Zurek phenomena \cite{Dziarmaga2005,Francuz2016}, discrete time-crystals \cite{Yang2021}, 
many-body localization \cite{Abanin2019} and the breaking of ergodicity \cite{Mishra2013,Chanda2016,Awasthi2018}.

In recent years, the concept of dynamical quantum phase transitions (DQPTs) have been introduced as nonequilibrium 
counterparts of thermal phase transitions \cite{Heyl2013,Heyl2018}. Within DQPTs real time plays the role of control 
parameter analogous to temperature in equilibrium phase transitions \cite{Jafari2019quench,Jafari2017,Najafi2019,Mukherjee2019,Zhang2016b,Serbyn2017,
Sadrzadeh2021,Wong2022,Rylands2021,Abdi2019}. 
While the conventional equilibrium phase transition is characterized by nonanalyticities in the thermal free energy, 
the DQPT is represented by the nonanalytical behavior of dynamical free energy 
\cite{andraschko2014dynamical,vajna2015topological,Karrasch2013,vajna2014disentangling,jafari2019dynamical,Mondal2022,Mendoza2022,
Sedlmayr2018,Sedlmayr2018b,Khatun2019,Ding2020,Corps2023,Nicola2021,Verga2023,Rossi2022,Khan2023anomalous,Khan2023}.
DQPT displays a phase transition between dynamically emerging quantum phases, that takes place during the nonequilibrium
coherent quantum time evolution under sudden/ramped quench 
\cite{Zhou2021,Vanhala2023,Mondal2023,Cao2020,Sedlmayr2023,Sedlmayr2022,Corps2022,Sedlmayr2020,
Zeng2023,Stumper2022,Yu2021,Vijayan2023,Zheng2023,Xue2023,Bhattacharjee2023,Leela2022,Puskarov2016,Zamani2024,Gesualdo2022} 
or time-periodic modulation of Hamiltonian 
\cite{yang2019floquet,Zamani2020,kosior2018dynamical,Jafari2021,kosior2018dynamicalb,Naji2022,Jafari2022,Naji2022b}. 
Furthermore, analogous to order parameters at equilibrium quantum phase transition, a dynamical topological order parameter is 
proposed to capture DQPTs \cite{budich2016dynamical,Bhattacharya}.

DQPT was observed experimentally in several studies 
\cite{flaschner2018observation,jurcevic2017direct,martinez2016real,guo2019observation,wang2019simulating,Nie2020,Francisco2022} 
to confirm theoretical anticipation. Most of these researches associated with deterministic quantum evolution generated by ramping or a sudden 
quench of the Hamiltonian. However, relatively little attention has been devoted to the stochastic driving of thermally isolated 
systems with noisy Hamiltonian. In any real experiment, the simulation of the desired time dependent Hamiltonian is imperfect and noisy 
fluctuations are inevitable \cite{Pichler,Zoller1981,Chen2010,Doria2011}. 
Therefore, understanding the effects of noise in such systems is of utmost importance both in designing experiments 
and comprehend the results \cite{Marino2012,Marino2014,Rahmani2016,Bando2020,Chenu2017}. 

Despite numerous studies of DQPTs in a wide variety of long-range quantum systems 
\cite{Dutta2017,Jafari2020,Xavier2023,Syed2021,Bojan2018,Mishra2020,Lakkaraju2023}, comparatively little attention has been paid 
to the noise effects \cite{Jafari2024} on long-range interaction properties. In the present work, we contribute to develop 
the systematic understanding of the competition between noise and long-range interaction at noisy ramped quench DQPT. For this purpose, 
we investigate the ramped quench DQPT of long-range pairing Kitaev model \cite{Vodola2014} in the presence of 
the white noise with Gaussian distribution \cite{Rahmani2016}. We solve an exact master equation for the quench dynamics averaged over the noise distribution.
This allows us to study the competition between the near-adiabatic quench dynamics of the gapped
modes of the long-range pairing system and the accumulation of noise induced excitations.

We show that, for the quench across a single critical point, while the long-range pairing induces a region with three DQPTs time scales 
(three critical modes), this region shrinks in the presence of the noise. 
In addition, for a quench that crosses two critical points, the critical sweep velocity above which the DQPTs disappear, 
enhances by the long-range pairing while decreases in the presence of the noise. In other words, the noise has destructive effects 
on long-range pairing features.

The paper is organized as follows. In Sec. \ref{DPT}, the dynamical free energy and DTOP of the two band Hamiltonians are discussed.
In Sec. \ref{model}, we present the model and review its exact solution and equilibrium phase transition.
Section \ref{results} is dedicated to the numerical simulation of the noiseless case based on the analytical result.
The effects of noise on the system is numerically studied in Section \ref{Noise}. Section \ref{SD} contains some concluding remarks.

\section{Quench of an integrable model and dynamical phase transition\label{DPT}}

\subsection{Dynamical free energy\label{DFE}}
%
\begin{figure}
\centerline{\includegraphics[width=1\linewidth]{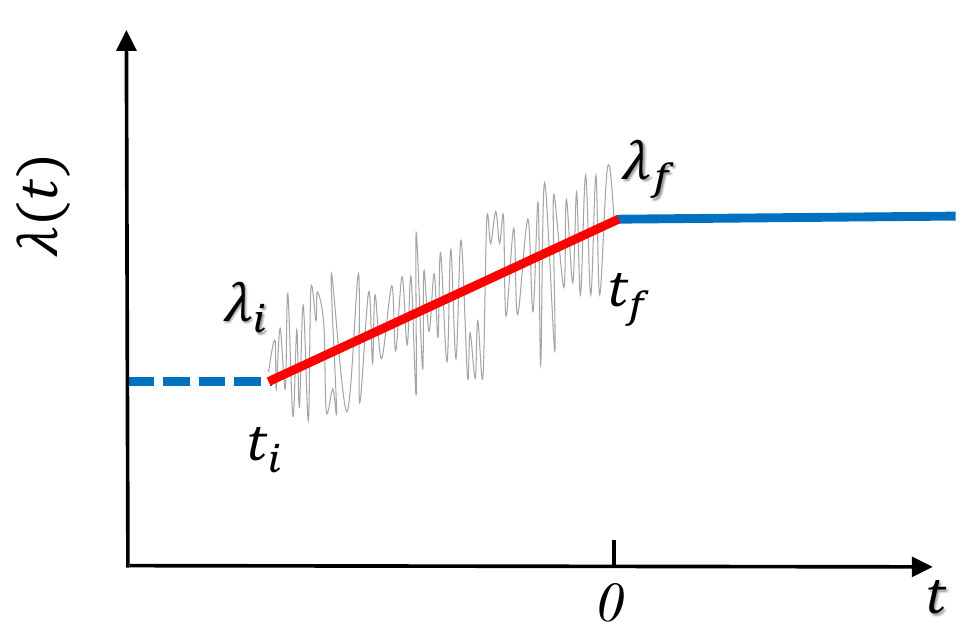}}
\caption{(Color online) Illustration of a linear ramped quench (red color).
Here, $\lambda(t)$ is the time-dependent parameter in Hamiltonian, $\lambda_i$ and $\lambda_f$ its initial and final values, 
and $t_i$ and $t_f=0$ the corresponding times. The wavy gray oscillations exhibit the presence of noise.}
\label{fig1}
\end{figure}
%

%
To study the ramped quench DQPTs, we follow the method used in Refs. \cite{Divakaran2016,Sharma2016b} in the subsequent discussions.
Let us consider an integrable model reducible to a two level Hamiltonian $H_k(\lambda)$ for each momentum mode and
the system is initially ($t_i \to-\infty$) prepared in the ground state $|g^i_k\rangle$ of the pre-quench Hamiltonian $H_k(\lambda_i)$ 
for each mode. Thereupon the parameter $\lambda$ is quenched from an initial value $\lambda_i$ at $t_i$ to the final value $\lambda_f$ at $t_f$, 
following the linear quenching protocol $\lambda(t)=vt$, in such a way that the system crosses the quantum critical point (QCP) at $\lambda = \lambda_c$. 
Since the adiabatic dynamics breaks in the vicinity of the QCP, the final state $|\psi^{f}_{k} \rangle$ (for the $k$-th mode) may not
be the ground state of the post-quench Hamiltonian $H_{k}(\lambda_f)=H^{f}_{k}$. The post-quench state can be written in the form of 
$|\psi^{f}_{k} \rangle = v_k |g_k^f\rangle+ u_k |e_k^f\rangle$, ($|u_k|^2 + |v_k|^2 =1$) where, $|g_k^f\rangle$ and $|e_k^f\rangle$ 
are the ground and the excited states of the post-quench Hamiltonian $H^{f}_{k}$, respectively with the corresponding energy eigenvalues 
{$\epsilon_{k,1}^f$ and $\epsilon_{k,2}^f$}.
The non-adiabatic transition probability where the system ends up in the excited state at the end of quench is denoted 
by $p_k=|u_k|^2=|\langle e_k^f|g^i_k\rangle|^2$. 
Therefore, the Loschmidt overlap and the corresponding dynamical free energy \cite{Heyl2013,Heyl2018}, for the mode $k$ 
for $t>t_f$ are defined by \cite{Divakaran2016,Sharma2016b}
%
\bea
\label{eq1}
{\cal L}_k&=&\langle \psi^{f}_{k} |\exp(-i H^{f}_{k} t)| \psi^{f}_{k}\rangle\\
\no
&=&|v_k|^2\exp(-i \epsilon_{k,1}^f t) + |u_k|^2 \exp(-i \epsilon_{k,2}^f t),\\
\label{eq2}
g_k(t) &=& -\frac{1}{N}\log \langle \psi^{f}_{k} |\exp(-i H^{f}_{k} t)| \psi^{f}_{k} \rangle
\eea
%
respectively, where $N$ is the size of the system.\\

Summing over the contributions from all modes and replacing summation by the integral in the thermodynamic limit, 
one gets \cite{Divakaran2016,Sharma2016b,Pollmann2013}
%
\bea
\label{eq2}
g(t)=\frac{-1}{2\pi} \int_{0}^{\pi} \ln \Big(1 + 4 p_k (p_k-1) \sin^2 (\frac{\epsilon_{k,2}^f-\epsilon_{k,1}^f}{2}) t \Big) dk,
\nonumber \\
\eea
where $t$ is measured from the instant the final state, $|\psi^{f}_{k}\rangle$, is reached
at the end of the ramped quench (Fig. \ref{fig1}).
The non-analyticities in $g(t)$ appear at the values of the real time $t_n^*$s given by
%
\bea
t_n^{\ast} =  \frac{\pi} {(\epsilon_{k^{\ast},2}^f-\epsilon_{k^{\ast},1}^f) }  \left(2n+ 1 \right).
\label{eq3}
\eea
%
These are the critical times for the DQPTs, with $k^{\ast}$ the mode at which the argument of the logarithm 
in Eq.~(\ref{eq2}) vanishes for $|u_{k^{\ast}}|^2=p_{k^{\ast}}=1/2$.

For the case $\epsilon_{k,2}^f = -\epsilon_{k,1}^f  = \epsilon_k^f$, Eq. \eqref{eq3} is simplified to
%
\bea
t_n^{\ast}= t^{\ast}\left(n+\frac 1 {2}\right),~~~~ t^{\ast}=\frac{\pi} {\epsilon_{k^{\ast}}^f}.
\label{eq4}
\eea
%

\subsection{Dynamical Topological Order Parameter\label{DTOP}}

The dynamical topological order parameter is introduced
to represent the topological characteristic associated with DQPTs \cite{budich2016dynamical}.
The DTOP displays integer (quantized) values as a function of time and its unit magnitude jumps at the time of 
DQPTs reveal the topological aspect of DQPT \cite{budich2016dynamical,Bhattacharjee2018,sharma2014loschmidt}.

The dynamical topological order parameter is defined as \cite{budich2016dynamical}
%
\begin{eqnarray}
\label{eq5}
N_w(t)=\frac{1}{2\pi}\int_0^\pi\frac{\partial\phi^G(k,t)}{\partial k}\mathrm{d}k,
\end{eqnarray}
%
where the geometric phase $\phi^G(k,t)$ is extracted from the total phase $\phi(k,t)$ by subtracting the dynamical
phase $\phi^{D}(k,t)$: $\phi^G(k,t)=\phi(k,t)-\phi^{D}(k,t)$.
The total phase $\phi(k,t)$ is the phase factor of Loschmidt amplitude in its polar coordinate representation,
i.e., ${\cal L}_{k}(t)=|{\cal L}_{k}(t)|e^{i\phi(k,t)},$ and $\phi^{D}(k,t)=-\int_0^t \langle \psi_{k}^{f}(t')|H(k,t')|\psi_{k}^{f}(t')\rangle dt',$
in which $\phi(k,t)$ and $\phi^{D}(k,t)$, for the two level system can be calculated as follows \cite{Divakaran2016,Sharma2016b}

%
\bea
\no
\phi(k,t)= \tan^{-1} \Big(\frac{-|u_k|^2 \sin(2\epsilon_k^f t)}{|v_k|^2 + |u_k|^2 \cos (2\epsilon_k^f t)} \Big),
\eea
%
%
\bea
\no
\phi^{D}(k,t) = -2 |u_k|^2 \epsilon_k^f t,
\eea
%
so that \cite{Divakaran2016,Sharma2016b}
%
\bea
\label{eq6}
\phi_k^G = \tan^{-1} \Big(\frac{-|u_k|^2 \sin(2\epsilon_k^f t)}
{|v_k|^2 + |u_k|^2 \cos (2\epsilon_k^f t)} \Big)  + 2 |u_k|^2 \epsilon_k^f t.
\eea
%
In the following we will study the DQPTs in the long-range pairing Kitaev model following the noiseless and noisy ramped quench 
and the corresponding topological properties (DTOP).

%
\begin{figure}
\centerline{\includegraphics[width=1\linewidth]{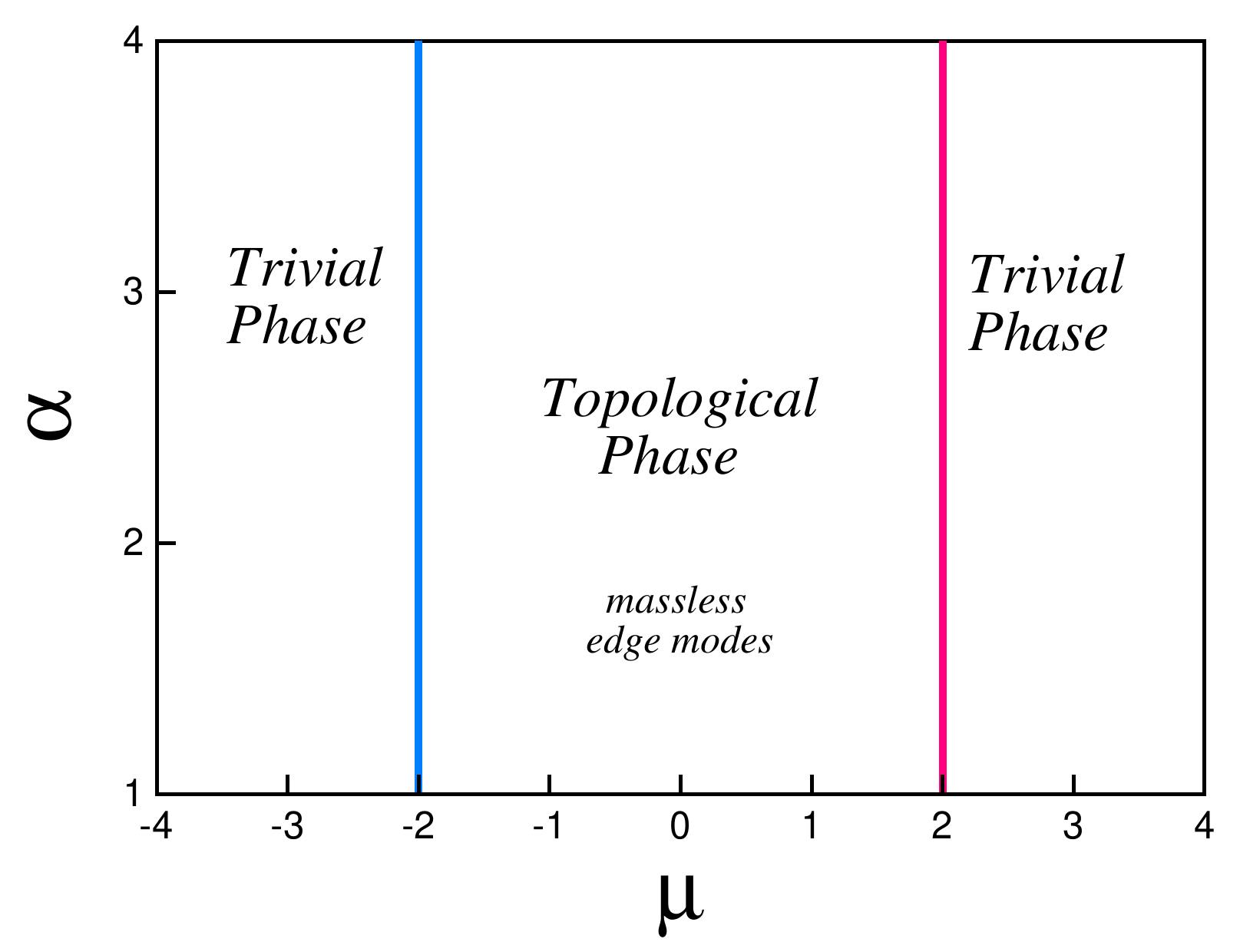}}
\caption{(Color online) Phase diagram of the long-range pairing Kitaev chain in the $\alpha-\mu$ plane
for $\alpha>1$.}
\label{fig2}
\end{figure}
%
\section{Model and Exact Solution\label{model}}

Recently, an extension of the Kitaev model \cite{Kitaev2001} which describes the algebraic 
decay of the tunneling and/or pairing terms has been intensively investigated \cite{Ryu2002,Vodola2014,Simon2020,Alecce2017,solfanelli2023}. 
This model describes experimental realizations of long-range topological superconductors \cite{Stevan2014,ruby2017}. 
It has been shown that the phase diagram is modified in the presence of the long-range interactions \cite{Vodola2014,Alecce2017}.  
Moreover, this model exhibits algebraically localized edge states and an algebraic closing of the energy gap \cite{Vodola2014,Alecce2017}. 
However, when the pairing and tunneling terms are isotropic, exponential localization is recovered independent of 
the power-law exponent, as long as it is larger than unity \cite{Alecce2017,Simon2020}.

In this paper, we investigate how the noise affects features of the long-range interaction in the long-range pairing Kitaev model. 
Representing fermionic annihilation (creation) operators as $c_n (c_n^{\dagger})$, the Hamiltonian of the long-range pairing Kitaev 
model with linear time dependent chemical potential is given as
%
\bea
\label{eq8}
H&=&-w\sum_{n=1}^{N}\left(c_n^{\dagger}c_{n+1}+h.c.\right)-\mu(t)\sum_{n=1}^{N}\left(c_n^{\dagger}c_{n}-{{1}\over{2}}\right)\nonumber\\
&&+{{\Delta}\over{2}}\sum_{n,\ell}d_\ell^{-\alpha}\left(c_n c_{n+\ell}+c_{n+\ell}^{\dagger}c_{n}^{\dagger}\right)
\eea
%
where $w$ denotes the hopping strength of the fermionic particles between adjacent lattice sites, 
$\Delta$ is the strength of the superconducting pairing term that decays with distance $l$ in a power law 
fashion characterized by exponent $\alpha$, and the onsite time dependent chemical potential $\mu(t)=\mu_f+vt$ changes from 
the initial value $\mu_i$ at time $t=t_i<0$ to the final values $\mu_f$ at $t=t_f=0$
with sweep velocity $v$. The effective distance $d_\ell$, between two sites denoted by $n$ and $n+\ell$ on the closed ring with $N$ sites,  
is given by the function $d_\ell= \min(\ell,N-\ell)$.

In the presence of the long-range pairing, the Hamiltonian Eq. (\ref{eq8}) is exactly solvable
in the momentum space \cite{Vodola2014}. Introducing the Nambu spinor $\Gamma^{\dagger}_{k_m}=(c_{k_m}^{\dagger},~c_{-k_m})$, 
the Fourier transformed Hamiltonian can be expressed as the sum of independent terms acting in the two-dimensional 
Hilbert space generated by $k$
%
\begin{eqnarray}
{\cal H}(t)=\frac{1}{2}\sum_{m=1}^{N/2}\Gamma_{k_m}^{\dagger}H^{(0)}_{k_m}(t) \Gamma_{k_m}
\label{eq9},
\end{eqnarray}
%
where $H^{(0)}_{k_m}(t)$ (the superscript in $H^{(0)}_{k_m}(t)$ is introduced to denote noise-free driving) is given by
%
\begin{equation}
\begin{aligned}
H^{(0)}_{k_m}(t) & = 
\begin{pmatrix}
-(2w \cos k_m+\mu(t)) & i \Delta f_{\alpha}(k_m) \\
-i \Delta f_{\alpha}(k_m) & (2w \cos k_m+\mu(t))
\end{pmatrix},
\end{aligned}
\end{equation}
%
where $f_{\alpha}(k_m)=\sum_{\ell=1}^{N-1}{{\sin (k_m \ell)}/{d_\ell^{\alpha}}}$ is the Fourier transform of the superconducting gap term
and $k_m=(2m-1)\pi/N,~~m=1,2,\cdots N/2$. In the thermodynamic limit $N\rightarrow\infty$, when $k_m$ gets continuous values, 
the function $f_{\alpha}(k_m)$, is described as 
$f^{\infty}_{\alpha}(k)=-\frac{i}{2}\left(\mathbf{Li}_{\alpha}(e^{i k})-\mathbf{Li}_{\alpha}(e^{-i k})\right)$ with 
$\mathbf{Li}_{\alpha}(z)=\sum_{\ell=1}^{\infty}{{z^\ell}/{\ell^{\alpha}}}$ being the polylogarithmic function of $z$ that 
vanishes in the limit $k\to 0$ and $k\to \pi$ for $\alpha>1$. When $\alpha<1$ the polylogarithmic function only vanishes in the limit 
$k\to \pi$.

In the limit of $\alpha\rightarrow\infty$, the model reduces to that of the short-range Kitaev chain with 
only nearest-neighbor pairing which is exactly solvable and its topological properties were unravelled by Kitaev \cite{Kitaev2001}. 
In this limit, for time-independent chemical potential $\mu(t)=\mu$ and $w=1$, the time-independent Hamiltonian undergoes 
topological quantum phase transitions at $\mu_c=\pm 2$, where the energy gap closes at $k=0,\pi$ \cite{Kitaev2001}.
For $\alpha>1$ the phase diagram and the topological properties of the long-range pairing Kitaev chain are identical to that of a
short-range Kitaev chain (Fig. (\ref{fig2})). However, as $\alpha$ approaches $1$, the bulk gradually starts becoming gapped
near $\mu=-2$ and for $\alpha<1$, $\mu=-2$ no longer remains a critical point \cite{Vodola2014}.

In the time dependent case $\mu(t)=\mu_f+vt$, the instantaneous eigenvalues and eigenvectors of 
time dependent Hamiltonian Eq.(\ref{eq8}), are given by
%
\bea
\label{eq11}
\varepsilon^{\pm}_{k_m}&=&\pm\varepsilon_{k_m}=\pm\sqrt{h^{2}_{z}(k_m,t)+h^{2}_{xy}(k_m)},\\
\no
|\chi^{-}_{k_m}(t)\rangle&=&\cos(\frac{\theta_{k_m}(t)}{2})|\uparrow\rangle-{\it i}\sin(\frac{\theta_{k_m}(t)}{2})|\downarrow\rangle,\\
\no
|\chi^{+}_{k_m}(t)\rangle&=&-{\it i}\sin(\frac{\theta_{k_m}(t)}{2})|\uparrow\rangle+\cos(\frac{\theta_{k_m}(t)}{2})|\downarrow\rangle,
\eea
%
where, $$\cos(\frac{\theta_{k_m}(t)}{2})=\frac{\varepsilon_{k_m}-h_z({k_m},t)}{\sqrt{2\varepsilon_{k_m}(\varepsilon_{{k_m}}-h_z({k_m},t))}},$$ $$\sin(\frac{\theta_{k_m}(t)}{2})=\frac{h_{xy}(k_m)}{\sqrt{2\varepsilon_{{k_m}}(\varepsilon_{{k_m}}-h_z({k_m},t))}},$$ 
with $h_{xy}({k_m}) =\Delta f_{\alpha}(k_m)$, and $h_{z}({k_m},t)=2w\cos({k_m})+\mu(t)$, and $|\chi^{\pm}_{k_m,t}\rangle$ 
are the adiabatic basis  of the system.\\

In such a case, if the system is prepared in its ground state at $t_i\rightarrow-\infty$ ($\mu_i\ll \mu_c=-2$), 
the probability that the $k$:th mode is found in the upper level at $t$ is given as (see Appendix {\ref{APA}}) 

%
\bea
\no
p_k=e^{-\pi\gamma^{2}/4}|U_{22}\cos(\frac{\theta_{k_{m}}(t)}{2})-\frac{\gamma e^{-{\it i}\pi/4}}{\sqrt{2}}
U_{12}\sin(\frac{\theta_{k_m}(t)}{2})|^2,\\
\label{eq12}
\eea
%
with $U_{22}=D_{\nu}(x),~~U_{12}=D_{\nu-1}(x),$
%
%
where, $D_{\nu}(x)$ is the parabolic cylinder function \cite{szego1954,abramowitz1988}, 
$\gamma=\Delta f_{\alpha}(k_m)/\sqrt{2v}$, $\nu={\it i}\gamma^{2}/2$, $x=2e^{{\it i}3\pi/4}\sqrt{v}\tau_{k}$, 
and $\tau_{k}=((\mu_f+vt)/2+w\cos(k))/v$.

\begin{figure*}
\begin{minipage}{\linewidth}
\centerline{\includegraphics[width=0.33\linewidth]{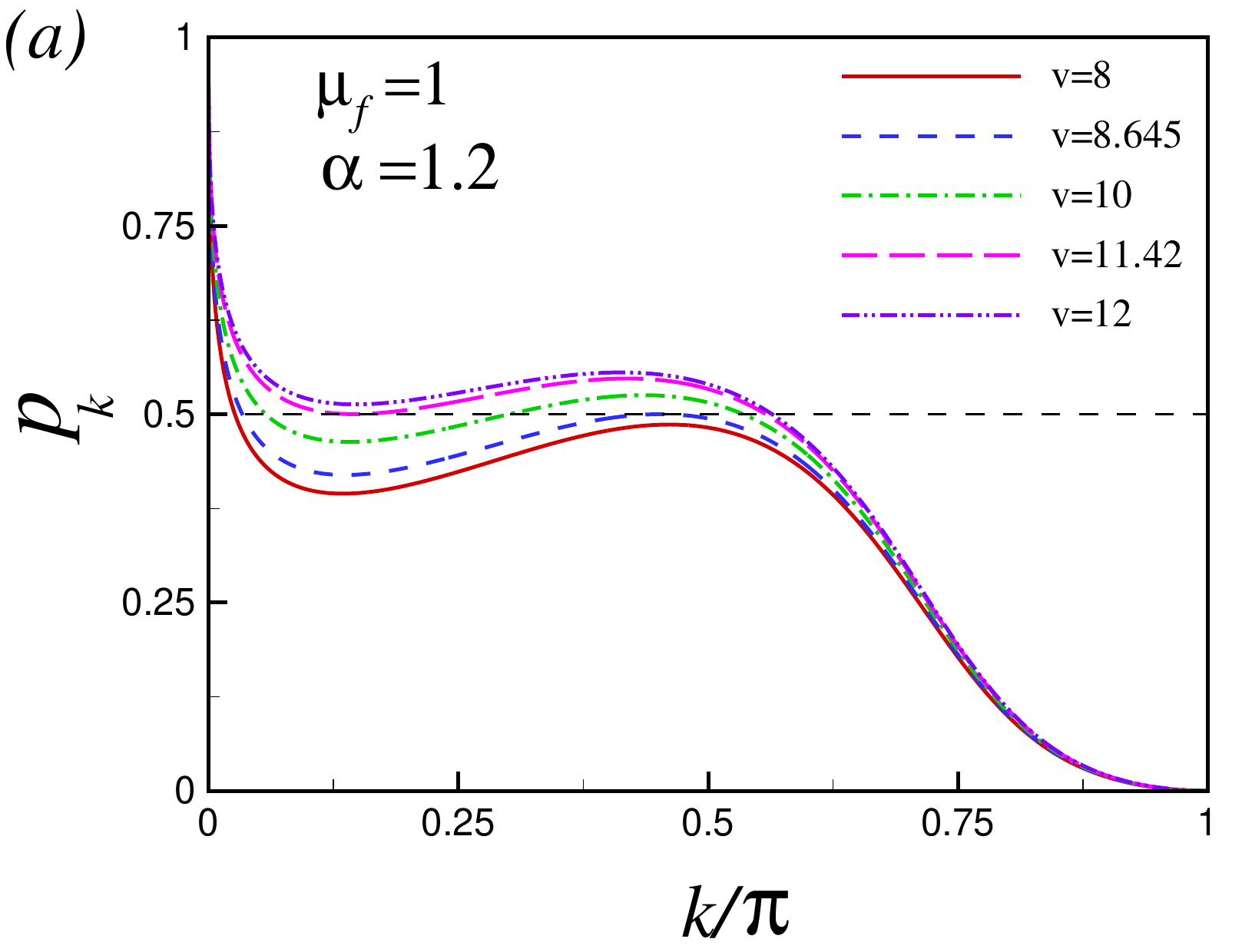}
\includegraphics[width=0.33\linewidth]{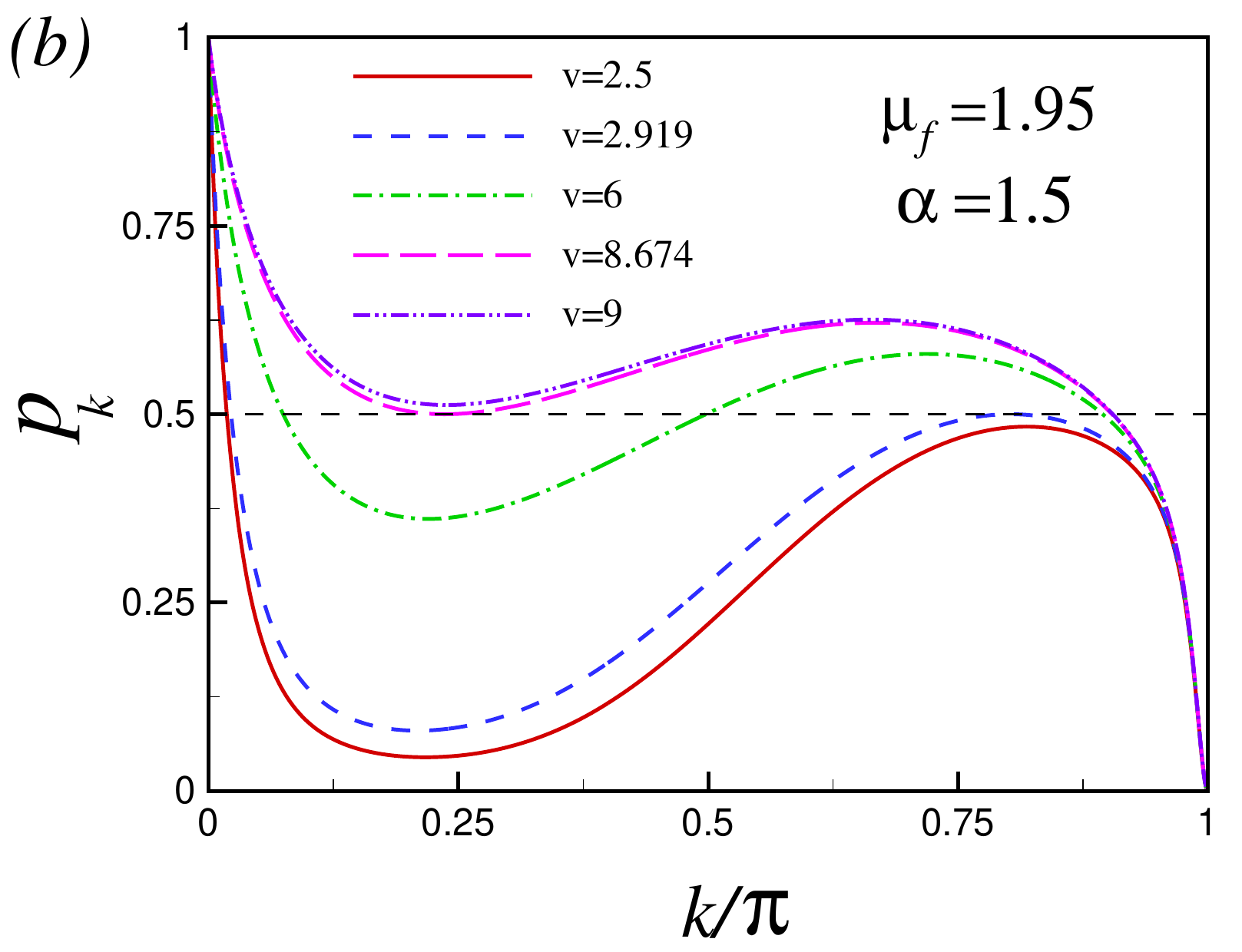}
\includegraphics[width=0.33\linewidth]{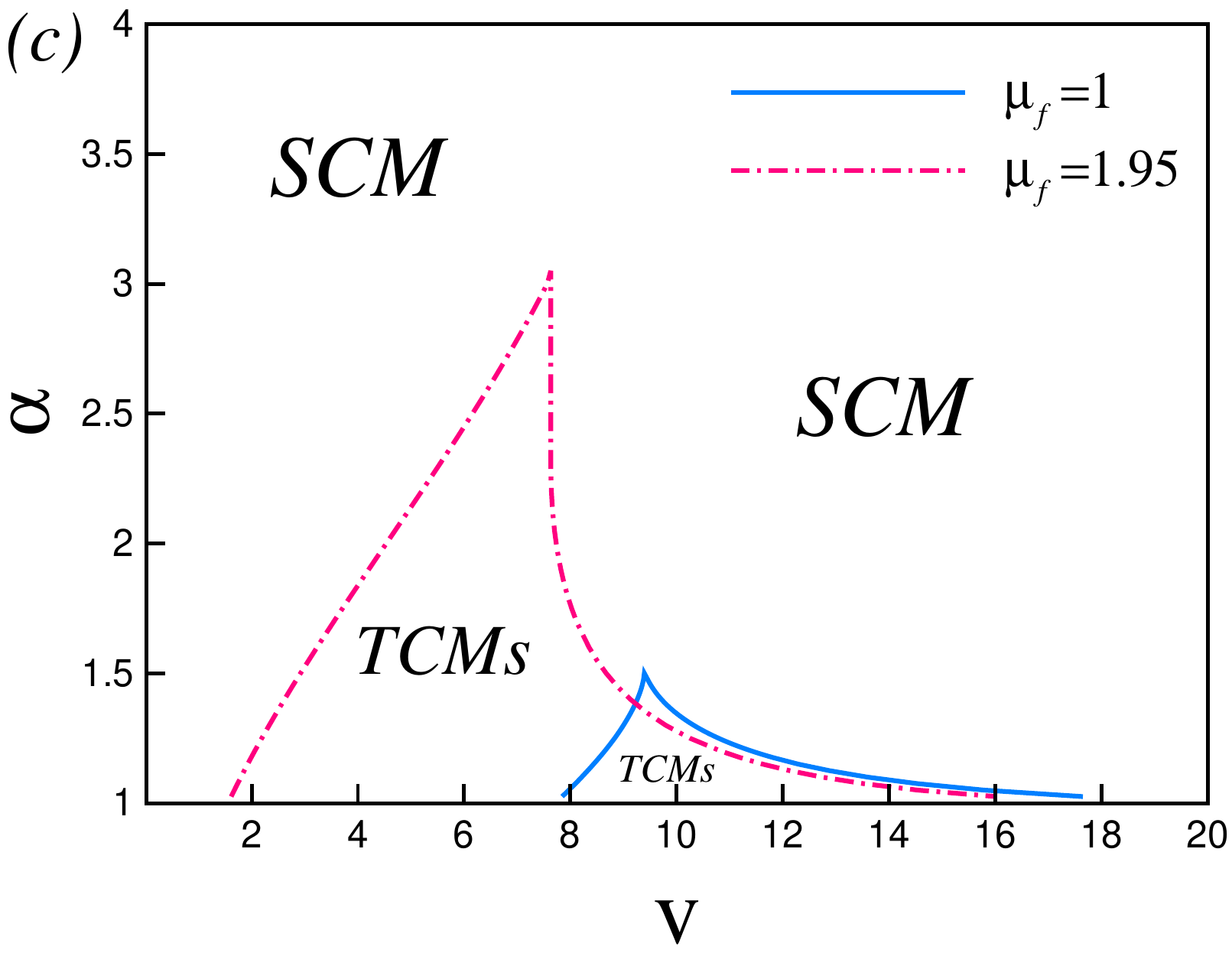}}
\centering
\end{minipage}
\begin{minipage}{\linewidth}
\centerline{\includegraphics[width=0.33\linewidth]{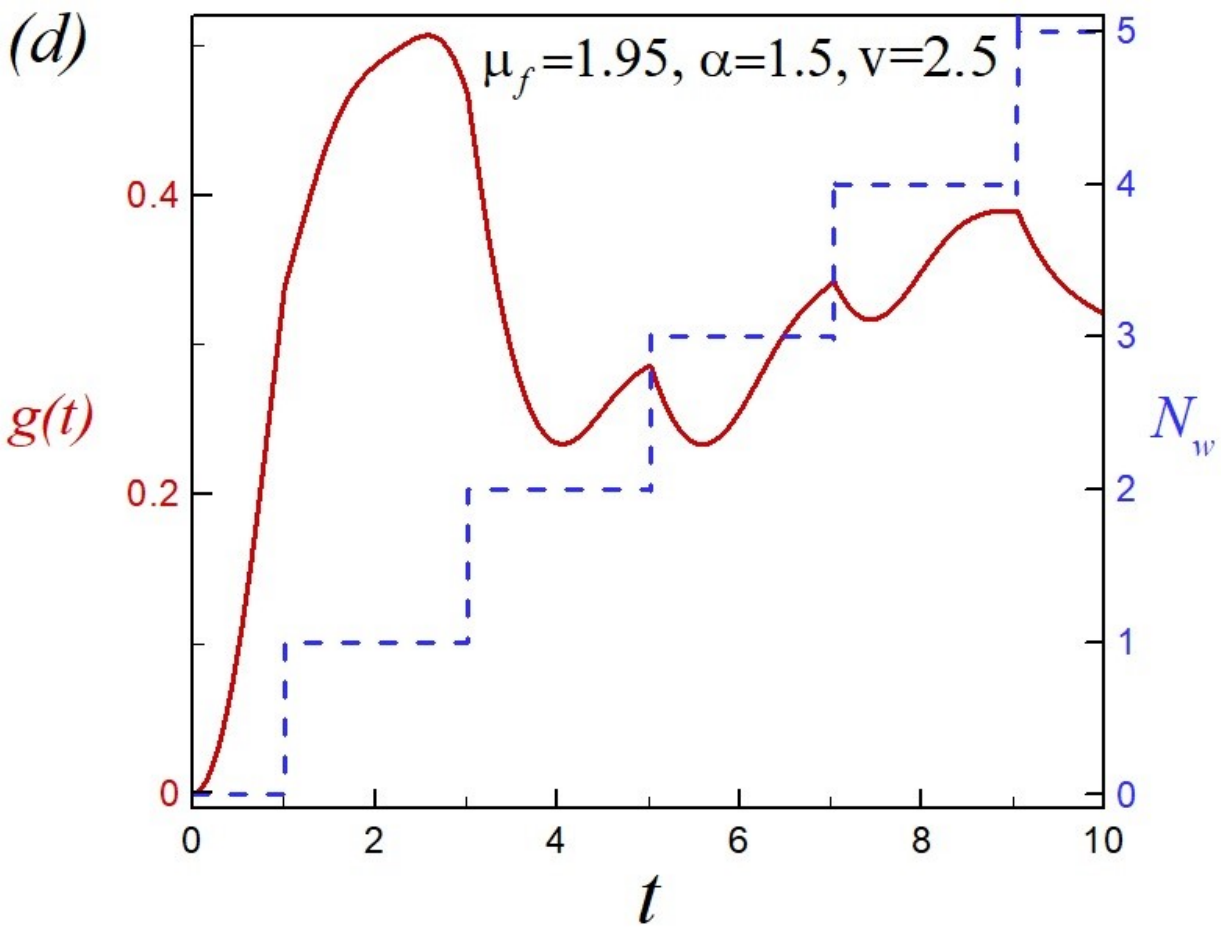}
\includegraphics[width=0.33\linewidth]{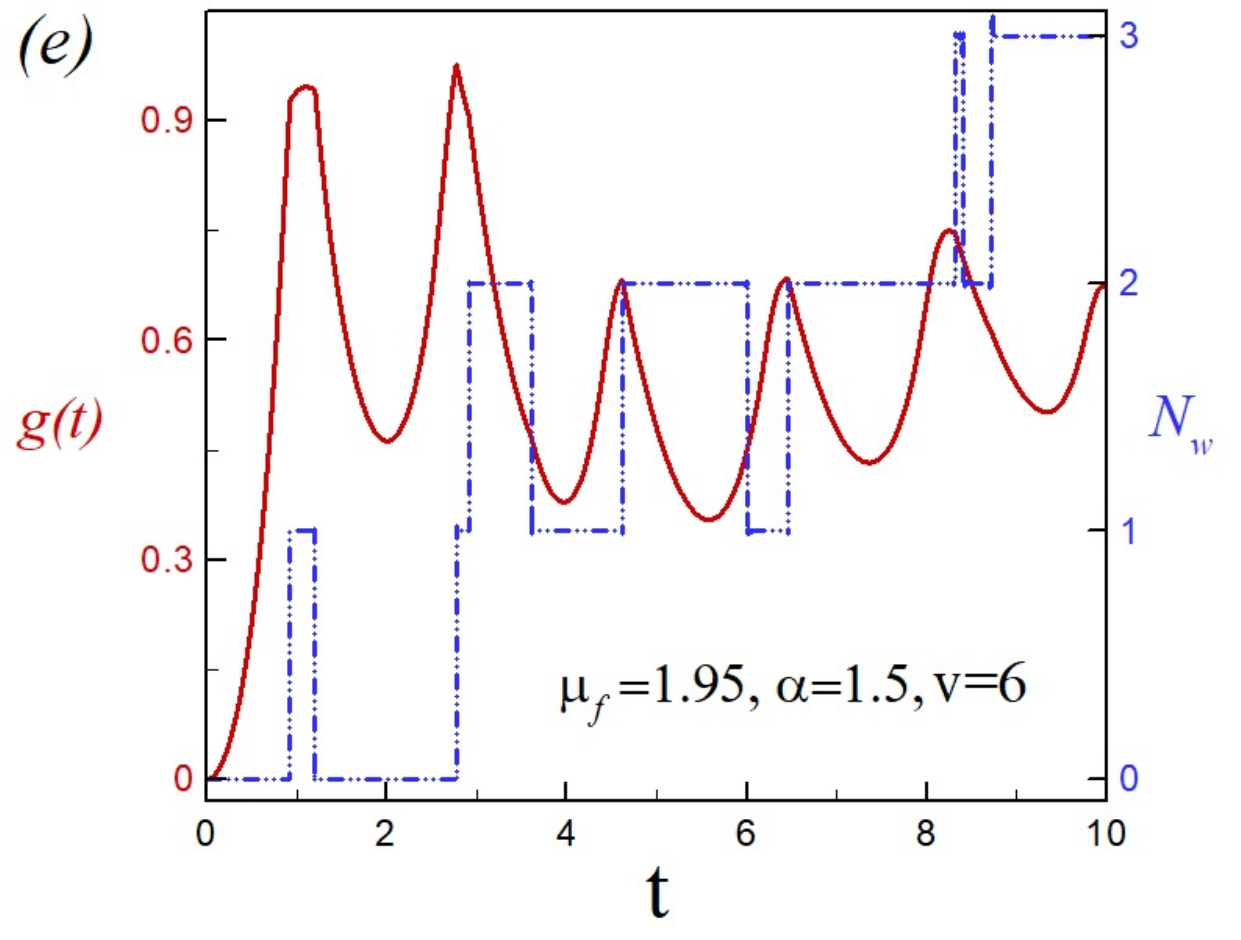}
\includegraphics[width=0.33\linewidth]{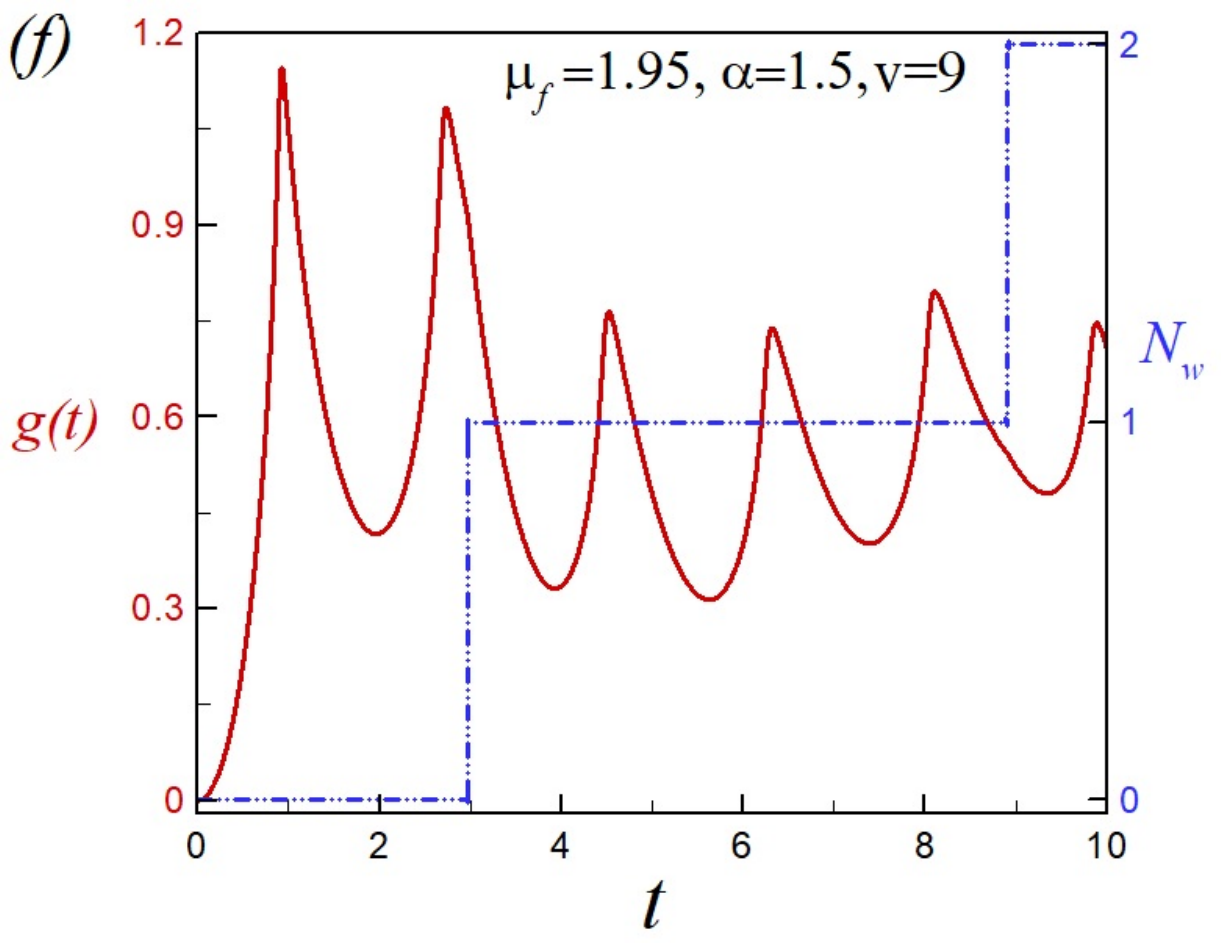}}
\centering
\end{minipage}
\caption{(Color online) Probability $p_k$ for finding the system with
momentum $k$ in the upper level for the noiseless ramped across the single critical 
point $\mu_c=-2$, for different sweep velocities, (a) for $\alpha=1.2$ and
$\mu_f=1$, (b) for $\alpha=1.5$ and $\mu_f=1.95$.
(c) The phase diagram of the model in $\alpha-v$ plane for a noiseless quench that crosses
the single critical point $\mu_c=-2$ for $\mu_f=1$ (solid line) and
$\mu_f=1.95$ (dashed-dotted line). The dynamical free energy $g(t)$ and its associated dynamical 
topological order parameter $N_w(t)$  for a noiseless quench across the single
critical point corresponding to Fig. \ref{fig3}(b) for (d) $v=2.5$, 
(e) $v=6$ and (f) $v=9$.}
\label{fig3}
\end{figure*}
%

\section{Noiseless numerical results\label{results}}
In this section, we report the results of our numerical simulations, based on an analytical approach, to
investigate the dynamics of the model using the notion of DQPTs.
To this end, we consider the linear quenching of the chemical potential $\mu(t) = \mu_f+vt$, changes 
from initial value $\mu_i\rightarrow -\infty$, where the system is prepared in its ground state, 
to various final 
values $\mu_f=1, 1.95, 4$ at $t_f=0$.
In addition, to better understand the effect of long-range pairing on the dynamics of system after 
the ramp quench, we will focus only on the case $\alpha>1$ where the location of the critical points in the parameter 
space is not altered by varying $\alpha$.

\subsubsection{Quench across a single critical point}

For the ramped quench, which crosses the single critical point $\mu_c=-2$ at $k=0$, the excitation probability 
after quench is $k$ dependent. As expected, when the system is driven across the critical point, 
the system undergoes nonadiabatic evolution due to the gap closing and thus the transition probability is 
maximum  at the gap closing mode $k=0$. However, away from the gap closing mode the system evolves 
adiabatically due to the non-zero energy gap and can be shown that $p_{k\rightarrow\pi}\rightarrow 0$.
Considering these two limiting cases, and also continuity of the transition probability as a function of $k$ 
in the thermodynamic limit, imply that there should exist a critical mode $k^{\ast}$ at which $p_{k^{\ast}}=1/2$ 
and consequently DQPTs occur. The transition probability has been plotted versus $k$ in Fig. \ref{fig3}(a), (b) 
for $\mu_f=1, 1.95$ for different sweep velocities as the ramped quench crosses the single critical point $\mu_c=-2$.
Since the quench crosses the critical point, the excitation probability takes its maximum value $p_{k}=1$ at $k=0$,
while it is negligible away from the gap closing mode ($k\rightarrow\pi$).
From these observations, it is straightforward to conclude that there is always a critical momentum $k^{\ast}$ and 
hence those of $t_n^*$, related through Eq.~(\ref{eq5}). Interestingly, we observed that there exists a region in the
parameter space $v-\alpha$ where the system encompasses three distinct critical modes $k^{\ast}$ for which 
$p_{k^{\ast}}=1/2$, even though the system is quenched across a single QCP. In such a case, the system displays 
three different critical time scales $t^{\ast}$ as obtained from Eq.~(\ref{eq5}).
While in the short-range case \cite{Zamani2024,Sharma2016b} the system contains only a single critical mode 
following a quench across a single QCP. In Fig.~\ref{fig3}(c), we have plotted a phase diagram in 
$v-\alpha$ plane for $\mu_f=1$, and $\mu_f=1.95$ in which region with three critical modes (TCMs) separated from the regions 
with single critical mode (SCM). On the phase boundary separating these two regions, there are two values of $k^{\ast}$ with $p_{k^{\ast}}=1/2$. 
As seen, the width of TCMs region shrinks and vanishes as $\alpha$ increases and also as $\mu_f$ decreases. 
The numerical results show that, the threshold values of $\mu_f$ above which TCMs region appears is $\mu_f\geq-0.1$.
In other words, the exponent $\alpha$ has a critical value $\alpha_c (v,\mu_f)$ above which the dynamical behavior of the system is similar 
to that of the short-range system. Consequently, our findings confirm that the appearance of TCMs region is indeed 
an artifact of the long-range pairing nature of the Hamiltonian.

The dynamical free energy $g(t)$ and DTOP ($N_w$) of the model have been depicted in Fig. \ref{fig3}(d)-(f) 
for the quench across a single critical point (corresponding to Fig. \ref{fig3}(b)), for different sweep velocities $v=2.5, 9$ and $v=6$, respectively.  
In Fig. \ref{fig3}(d) and (f) the system is in SCM region, where it encompasses a single critical time scale $t^{\ast}$.
Although the cusps in $g(t)$ are not discernible but the quantization and jumps in the associated DTOP are clearly 
visible as an indicator of DQPTs. 
{\color{black} The observed oscillation in the dynamical free energy seems to be the natural behavior, which results from the unitary time evolution of the post-quench ground state in terms of the Hamiltonian's eigenstates.}
The behaviour of DTOP, i.e, whether $N_w(t)$ would jump or drop, can be predicted by 
the slope of $p_k$ at the critical momentum $k^{\ast}$ (positive (negative) slop results jump (drop)) \cite{Sharma2016b,Dutta2017}. 
Appearing successive jumps or drops in $N_w(t)$ indicate that the system has only a single critical mode, while the presence of 
both jump and drop in DTOP curve implies that the system has at least two critical modes as seen in Fig. \ref{fig3}(e). As seen, uninterrupted jumps in Fig. \ref{fig3}(d), (f) 
reveal that the system is in SCM region while two successive jumps and then drop of $N_w$(t) in Fig. \ref{fig3}(e) points the existence 
of three different critical modes in accordance with $p_k$ (Fig. \ref{fig3}(b)), which shows that the system is in TCMs region.

%
\begin{figure*}
\begin{minipage}{\linewidth}
\centerline{\includegraphics[width=0.33\linewidth]{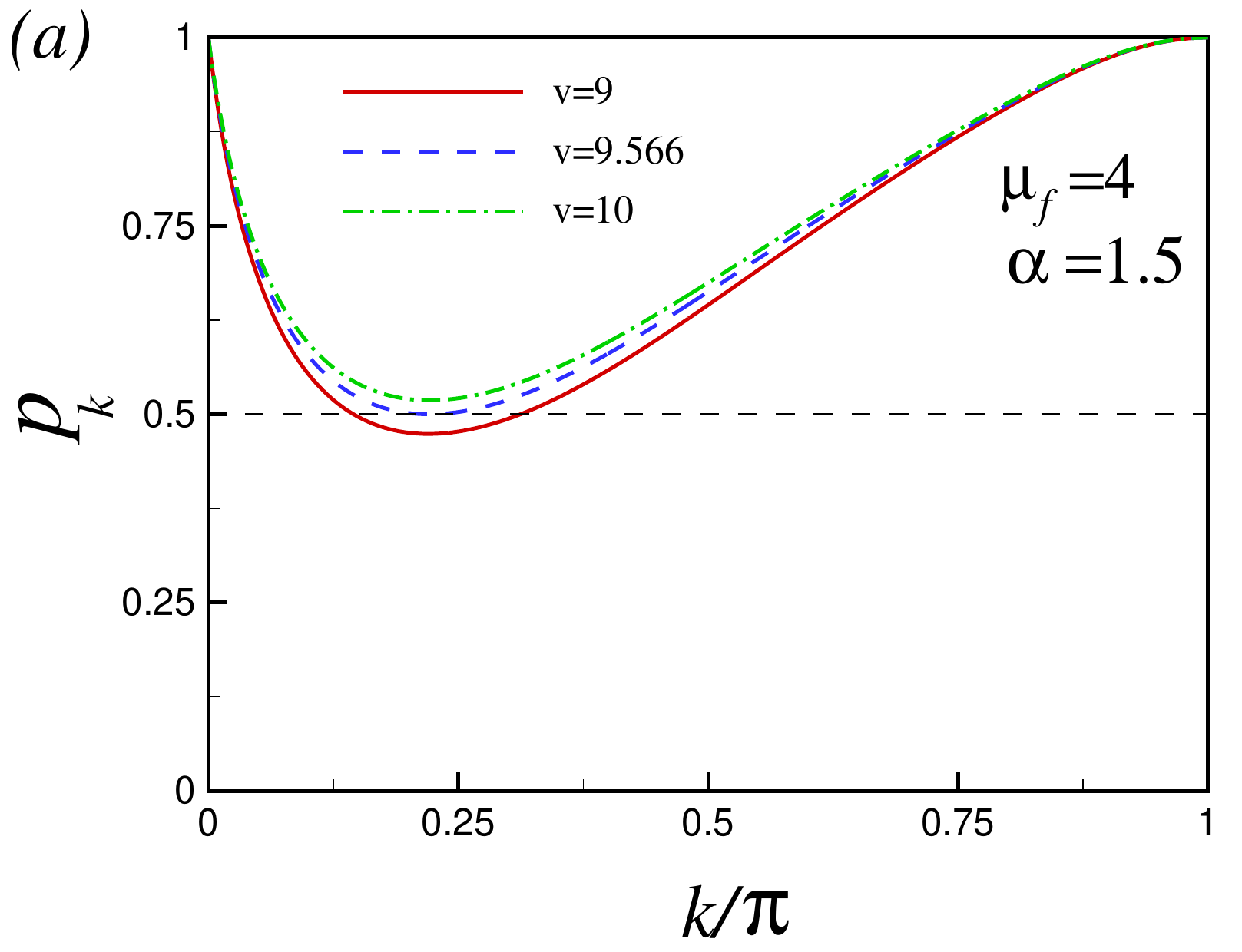}
\includegraphics[width=0.33\linewidth]{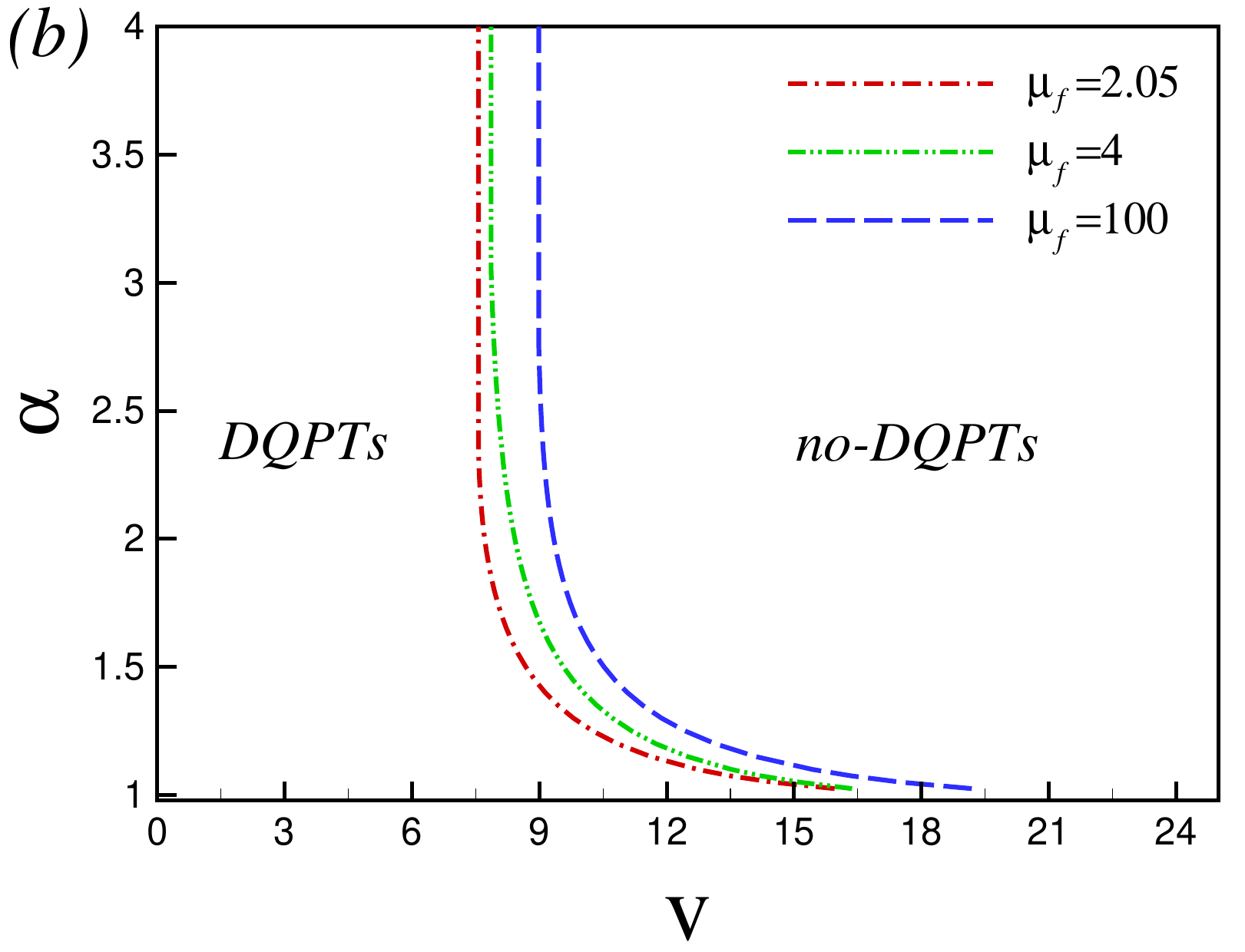}
\includegraphics[width=0.33\linewidth]{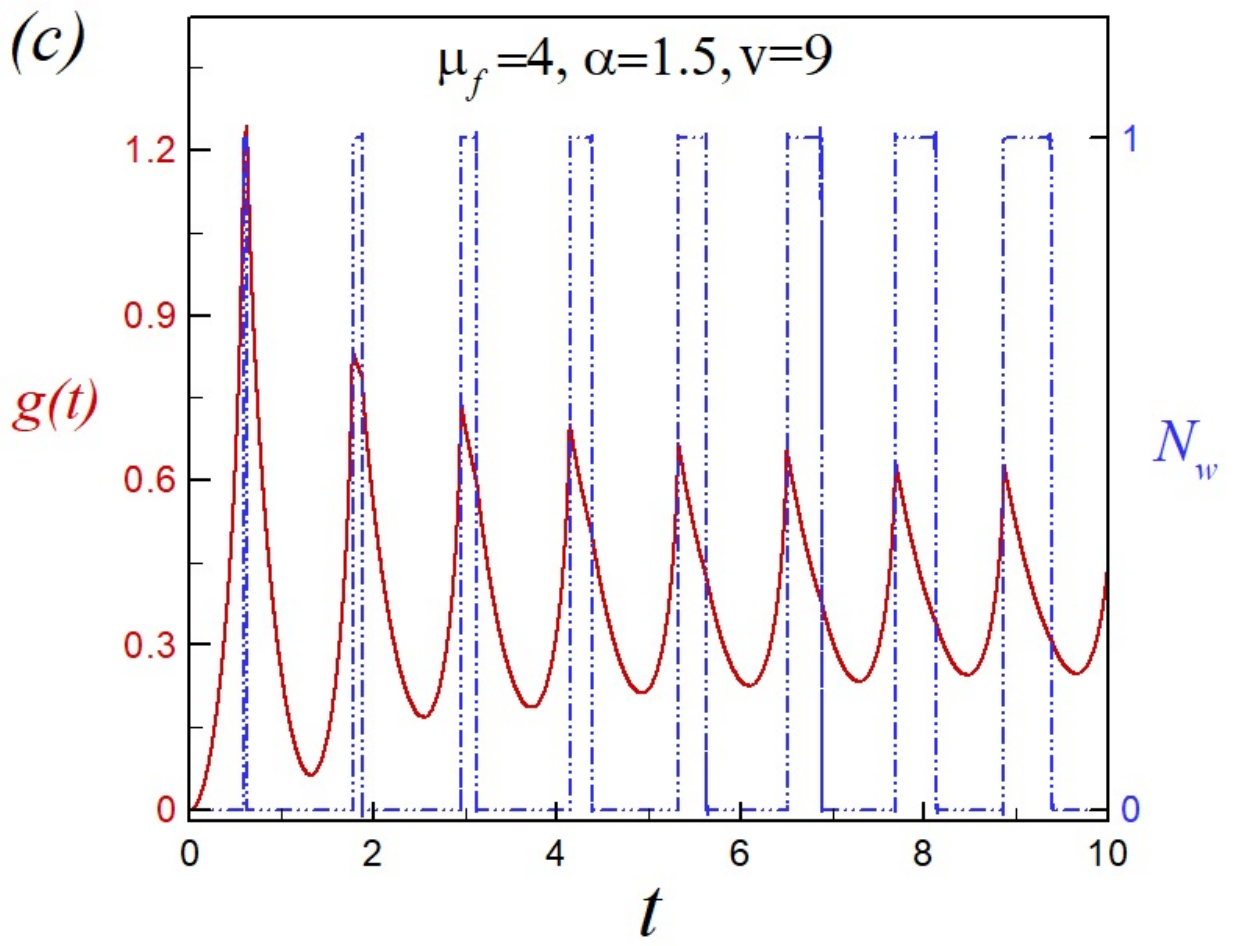}}
\centering
\end{minipage}
\caption{(Color online) (a) Probability of the excitation for noiseless ramped quench which crosses two 
critical points $\mu_c=\pm2$, for different sweep velocities, $\alpha=1.5$ and
$\mu_f=4$. (b) The phase diagram of the model in $\alpha-v$ plane for a noiseless quench 
that crosses two critical points $\mu_c=\pm2$ for $\mu_f=2.05$ (dashed-dotted line), 
$\mu_f=4$ (dashed-dotted-dotted line) and  $\mu_f=4$ (dashed line). 
(c) The dynamical free energy $g(t)$ and the associated dynamical topological 
order parameter $N_w(t)$  for a noiseless quench crosses two critical points corresponding to 
Fig. \ref{fig4}(a)  for $v=9$.}
\label{fig4}
\end{figure*}
%
\subsubsection{Quench across two critical points}
Performing a quench across both equilibrium critical points $\mu_c=\pm2$ shows new features. 
In these cases, the chemical potential is swept from one trivial (non-topological) phase to
another one, and it is not expected to result in DQPTs when the quench is sudden \cite{vajna2014disentangling,Sharma2016b,Pollmann2013}.   
For a quench crossing both critical points, as expected, the nonadiabatic evolution of the system at gap closing 
modes $k=0,\pi$, leads to maximum transition probability, i.e., $p_{k=0,\pi}=1$. 

However, the minimum of $p_k$, occurs at the maximum energy gap mode at $k=\pi/2$, which is the farthest mode from the gap closing mode.
Since, the maximum value of transition probability $p_{k=0,\pi}=1$ is greater than $1/2$, the appearance  
of DQPTs requires the condition that the minimum value of transition probability becomes less than $1/2$. 
As the system changes adiabatically at the gapped mode for small sweep velocity, making the quench sufficiently slow ($v < v_c$) 
ensures that the minimum excitation probability is smaller than $1/2$, which sets a succession of DQPTs.
In Fig. \ref{fig4}(a) the transition probability has been shown versus $k$ for a quench that crosses two critical points (i.e. $\mu_f=4$)
for the exponent $\alpha=1.5$. 
As predicted, $p_{k=0,\pi}=1$ and the minimum of $p_k$ away from the critical modes is less than $1/2$ for the small sweep velocity
($v<v_c=9.566$). In such a case, there is two critical modes $k^{\ast}_{\beta}$ and $k^{\ast}_{\gamma}$ at which $p_{k^{\ast}_{\beta,\gamma}}=1/2$ 
yields a sequence of DQPTs at the corresponding critical times $t^{\ast}_n=t^{\ast}_{n,\beta}, t^{\ast}_{n,\gamma}, n=0,1,\ldots$.
Furthermore, the minimum of $p_k$ becomes greater than $1/2$ for a sweep velocity greater than the critical sweep velocity $v=10>v_c=9.566$,
thus blocking the appearance of DQPTs.

The phase diagram of the model for a quench crossing two critical points, has been illustrated in Fig. \ref{fig4}(b) for
different values of $\mu_f=2.05, 4$ and $\mu_f=100$ where the region marked "DQPTs" support aperiodic sequences of DQPTs.
As seen, the critical sweep velocity $v_c$ decreases by increasing the exponent $\alpha$ and $v_c$ is equivalent to the 
that of short-range pairing system for $\alpha>2$.
 
Fig. \ref{fig4}(c) shows the dynamical free energy and DTOP for a quench crossing two critical points $\mu_c=-2$ 
and $\mu_f=2$. Cusps in $g(t)$ and quantizations in the associated DTOP are clearly visible as an indicator of DQPTs.
As observed,  DTOP oscillates between $0$ and $1$, which indicates that the corresponding $p_k$ contains two critical modes with different slopes (Fig. \ref{fig4}(a)).

\section{Noisy ramp quench\label{Noise}}
As mentioned the noises are ubiquitous and indispensable in any physical system. Specifically, when energy is
transferred into or out of an otherwise isolated system via a quench in the laboratory, there will inevitably 
be time dependent fluctuations ("noise") in this transfer. In this section we investigate the effects of noise 
on the dynamical phase diagram of the long-range pairing Kitaev model.
For this purpose, we add a noise to the time dependent chemical potential $\mu(t)=\mu_f+vt+R(t)$,
where $R(t)$ is a random fluctuation confined to the ramp interval $[t_i,t_f=0[$, with vanishing mean, $\langle R({t})\rangle=0$.
We use white noise with Gaussian two-point correlations $\langle R(t)R(t')\rangle=\xi^2 \delta (t-t')$ where
$\xi$ characterizes the strength of the noise ($\xi^2$ has units of time).  
White noise is approximately equivalent to fast colored noise with exponentially decaying two-point correlations 
(Ornstein-Uhlenbeck process)~\cite{Jafari2024}. 
In the presence of noise the transition probability is obtained by numerically solving the exact master equation 
\cite{luczka1991quantum,Budini2000,Filho2017,Kiely2021} 
for the averaged density matrix $\rho_{k_m}(t)$ of the noisy system
%
\begin{eqnarray}
\no
{d\over dt}\rho_{k_m}(t)=-i[H^{(0)}_{k_m}(t),\rho_{k_m}(t)]-\frac{\xi^2}{2}[H_1,[H_1,\rho_{k_m}(t)]],\\
\label{eq:master}
\end{eqnarray}
%
where $H^{(0)}_{k_m}(t)$ is the noise-free Hamiltonian while $R(t) H_1= - R(t) \sigma^{z}$ expresses the ``noisy" part for the full Hamiltonian $H^{(\xi)}_{k_m}(t)=H^{(0)}_{k_m}(t)+R(t)H_1$.
The master equation, Eq.(\ref{eq:master}), is solved within the quench interval $t \in\,  [t_i,0[$. 

The transition probability $p_k$ in the presence of the noise is given by
$$p_{k_m}=\langle \chi^{+}_{k_m}(t_f)|\rho_{k_m}(t_f)|\chi^{+}_{k_m}(t_f)\rangle.$$

As a result, the dynamical phase diagram of the model is characterized by the interplay of two competing effects: 
(i) The non-trivial excitation  resulting from the long-range pairing and (ii) the accumulation of noise-induced 
excitations during the evolution. Moreover, we expect that the non-adiabaticity by large values of the sweep velocity 
gives less time for the noise to become effective. Our numerical simulation, which is based on the exact master equation reveals that 
the main effect of noise is to shift the critical mode yielding the succession of DQPTs and a shift on the phase boundaries. 
In addition, the numerical results uncover that the noise contributions diminish the long-range pairing dynamics. 

\begin{figure}
\centerline{\includegraphics[width=1\linewidth]{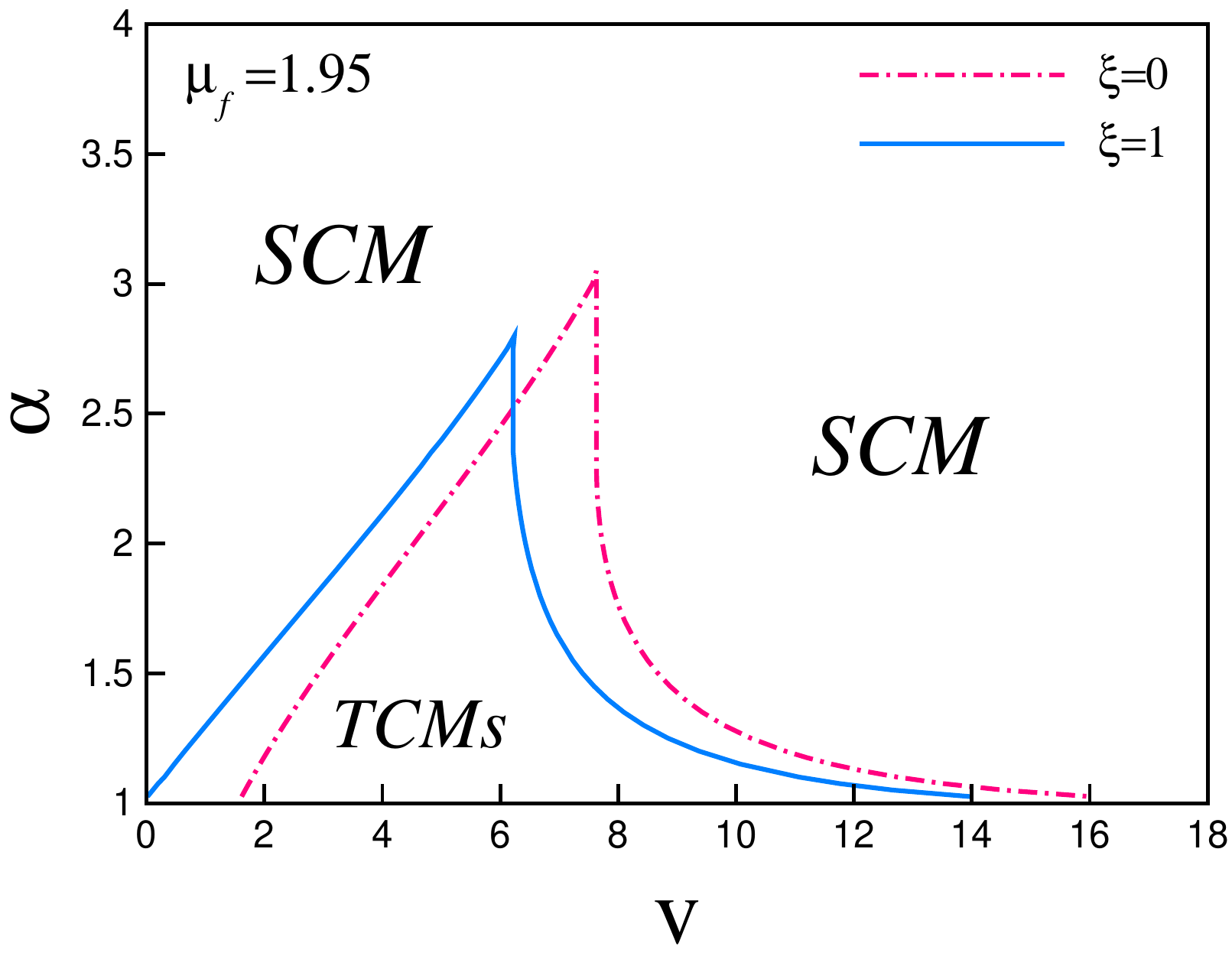}}
\caption{(Color online) The phase diagram of the underlying model in $\alpha-v$ plane for a noiseless
and noisy quench that crosses a single critical point $\mu_c=\-2$ 
for $\mu_f=1.95$. The dashed-dotted line represents the boundary between
TCMs and SCM region for the noiseless case and solid line displays
the boundaries for the noise intensity $\xi=1$.}
\label{fig5}
\end{figure}

The phase diagram of the model in the absence and presence of the noise ($\xi=0, 1$) has been plotted in Fig. \ref{fig5} 
for a quench across the single critical point for $\mu_f=1.95$. As seen, the TCMs boundaries change in the presence 
of the noise. Moreover, the width of TCMs region shrinks rapidly as exponent $\alpha$ increases.
In addition, the borders between TCMs and SCM regions change less for large 
values of sweep velocity which corresponds to our anticipation. In other words, the noise weakens the effect of 
long-range pairing on the dynamical phase diagram.
%
%

\begin{figure}[!t]
\centerline{\includegraphics[width=1\linewidth]{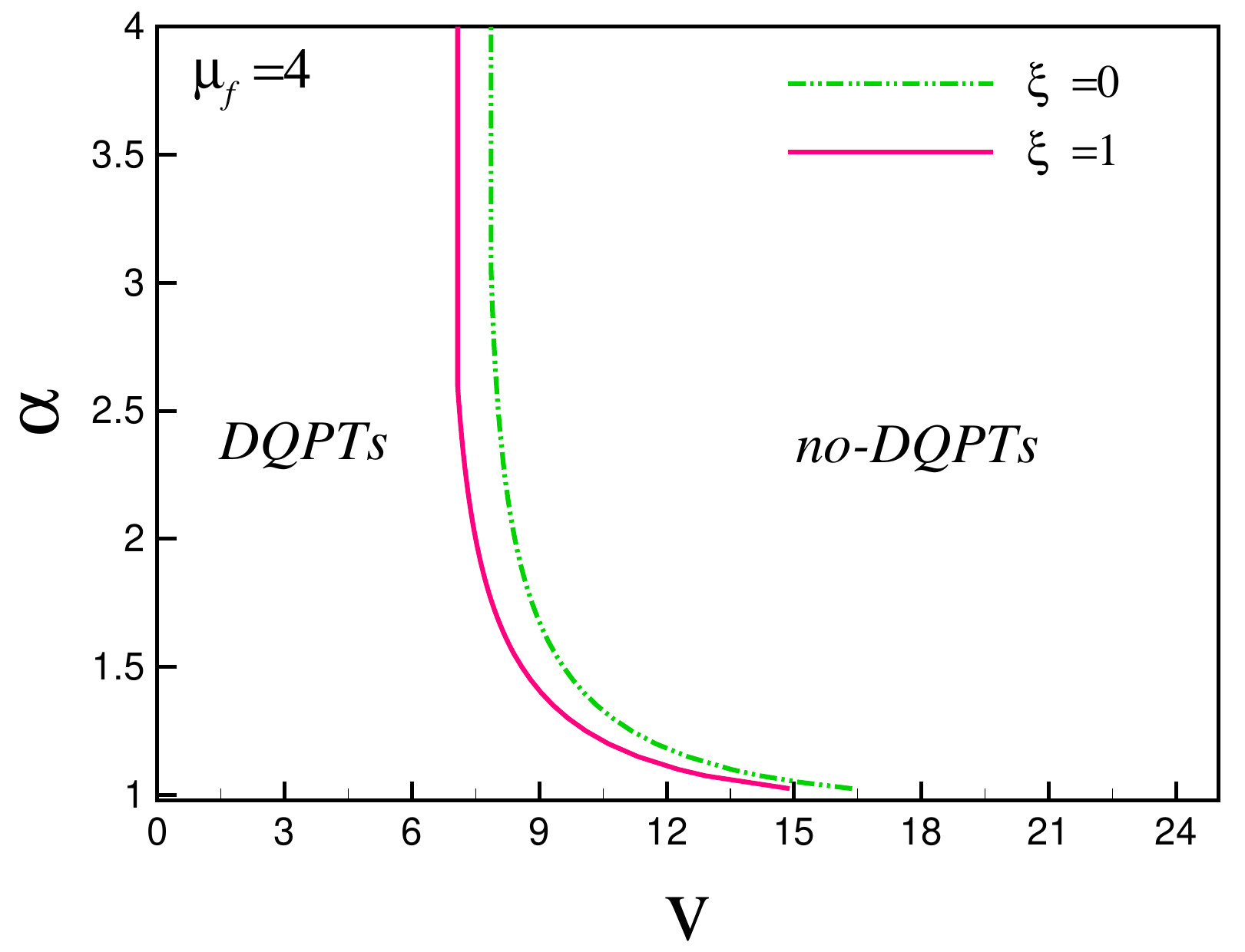}}
\caption{(Color online) The $\alpha-v$ phase diagram of the model for a noiseless and noisy quench which crosses two critical points $\mu_c=\pm2$ for $\mu_f=4$. 
The dashed-dotted-dotted line shows the boundary between
DQPTs and no-DQPTs regions for the noiseless case and solid 
line represent the corresponding border for the noise intensity $\xi=1$.}
\label{fig6}
\end{figure}

Fig. \ref{fig6} depicts the border between DQPTs and no-DQPTs regions for both noiseless $\xi=0$ and noisy $\xi=1$ cases, 
for a ramped quench that crosses two critical points for $\mu_f=4$. As indicated, the critical sweep velocity above which
the DQPT is wiped out, decreases in the presence of the noise even for large values of $\alpha$, where the critical values of 
the sweep velocity is the same as that of short range pairing case. Moreover, the border undergoes more changes for the smaller 
sweep velocities than the larger sweep velocities. However, the changes is constant for $\alpha>2$, where the dynamics is the same as that of 
the short-range pairing case. 
The numerical results show that the changes in the border of different regions in the phase diagram of both ramped quench cases
(Figs. \ref{fig5} and \ref{fig6}) decreases by decreasing the noise strength, as anticipated.

%
%
\section{Summary and discussion\label{SD}}
In this paper, we have studied the non-equilibrium dynamics of the long-range pairing Kitaev model ($\alpha > 1$) with 
noiseless and noisy linear time dependent chemical potential. 
For a noiseless quench across one of the equilibrium quantum critical points ($\mu_c=-2$), we find that
the dynamical phase diagram in $\alpha-v$ plane is classified into two regions, the three critical modes and the single critical mode regions.
The three critical modes region is the result of the long-range pairing,
in contrast to the short-range Kitaev model which shows a single critical mode for a noiseless quench across a single critical point. 
The three critical modes region shrinks and disappears as the exponent $\alpha$ increases. In addition,
the numerical results show that appearance of the three critical modes region depends on the final values of 
the chemical potential. The lower bond of the chemical potential above which three critical modes region 
emerges is $\mu_f=-0.1$ and the upper bound is the next critical point, i.e, $-0.1\leq\mu_f<2$.
Moreover, the exponent $\alpha$ has a critical value $\alpha_c (v,\mu_f)$ above which the dynamical behavior 
of the long-range pairing system is similar to that of the short-range system. Consequently, our finding confirms
that the appearance of TCMs region is indeed an outcome of the long-range pairing nature of the Hamiltonian.

Further, for a noiseless ramped quench that crosses two critical points, the critical sweep velocity above which the dynamical 
quantum phase transition is wiped out for long-range pairing, is larger than that of the short-range pairing case. 
The critical sweep velocity decreases by increasing the exponent $\alpha$ of long-range pairing and saturates to the critical sweep velocity
of the short-range pairing case beyond $\alpha=2$.

The boundaries between different regions in both cases of the ramped quench, are changed in the presence of the Guassian white noise. 
The three critical modes region for the quench that crosses the single critical point, shrinks faster in the presence of
noise by increasing $\alpha$. Moreover, for the ramped quench which crosses two critical points, the critical sweep velocity above which the
dynamical quantum phase transition disappears, reduces by adding noise. The numerical results exhibit that the system is affected less 
at the large sweep velocities. In summary, the noise has destructive effects on the long-range pairing features.
 
{\color{black} The case of $\alpha < 1$ hosts massive edge modes for the open boundary condition, which is not our case. However, the study of a system with massive edge modes could be an interesting issue for further investigations.}

\appendix

\section{Time-dependent Schr\"{o}dinger equation in the diabatic basis \label{APA}}

The time-dependent Schr\"{o}dinger equation of Hamiltonian in Eq. (\ref{eq8}) is given by
%
{\small
\bea
\no
{\it i}\frac{d}{dt}
\left(
 \begin{array}{c}
 a_1(t) \\
 a_2(t) \\
\end{array}
\right)
=
\left(
\begin{array}{cc}
-h_z(k,t) & i h_{x,y}\\
-i h_{x,y} & h_z(k,t)
\end{array}
\right)
\left(
 \begin{array}{c}
a_1(t) \\
a_2(t) \\
\end{array}
\right),\\
\label{eqAPA1}
\eea
}
%
where $h_z(k,t)=(2w \cos k_m+\mu(t))$, $h_{x,y}=i \Delta f_{\alpha}(k_m)$ and $a_1 (t)$, $a_2 (t)$
are the coefficients which define the quantum state in the diabatic bases.
The time-dependent Schr\"{o}dinger equation (\ref{eqAPA1}) is mapped to the time-dependent Schr\"{o}dinger equation
of Landau-Zener problem \cite{Vitanov1996,Vitanov1999} by performing $\pi/2$ rotation around the $z$ axes and defining the new 
time scale $\tau_{k}=(\mu_f+vt+2w\cos(k))/2v$,
%
{\small
\bea
\no
{\it i}\frac{d}{d\tau_{k_m}}
\left(
 \begin{array}{c}
a_1 (\tau_{k_m}) \\
a_2 (\tau_{k_m}) \\
\end{array}
\right)
=
\left(
\begin{array}{cc}
 -2v\tau_{k_m} & \Delta f_{\alpha}(k_m)  \\
\Delta f_{\alpha}(k_m) & 2v\tau_{k_m} \\
\end{array}
\right)
\left(
 \begin{array}{c}
a_1 (\tau_{k_m}) \\
a_2 (\tau_{k_m}) \\
\end{array}
\right).\\
\label{eqAPA2}
\eea
}
%

The Landau-Zener problem is exactly solvable as explained in Refs. \cite{Vitanov1999,Vitanov1996}
and the transition probability is given by Eq. ($\ref{eq12}$).

\bibliography{REF}

\end{document}